\documentclass[utf8]{FrontiersinHarvard} 

\usepackage{url,hyperref,lineno,microtype,subcaption}
\usepackage[onehalfspacing]{setspace}
\usepackage{siunitx}

\def\keyFont{\fontsize{8}{11}\helveticabold }
\def\firstAuthorLast{Marwan} 
\def\Authors{Norbert Marwan\,$^{1,2,*}$}
\def\Address{$^{1}$Potsdam Institute for Climate Impact Research (PIK), Member of the  Leibniz Association,
Telegrafenberg A31, 14473 Potsdam, Germany \\
$^{2}$University of Potsdam, Institute of Geoscience, Karl-Liebknecht-Straße 32, 14476 Potsdam, Germany}
\def\corrAuthor{Norbert Marwan}

\def\corrEmail{marwan@pik-potsdam.de}

\begin{document}
\onecolumn
\firstpage{1}

\title[Recurrence analyses of event time series]{Challenges and perspectives in recurrence analyses of event time series} 

\author[\firstAuthorLast ]{\Authors} 
\address{} 
\correspondance{} 

\extraAuth{}

\maketitle

\begin{abstract}
The analysis of event time series is in general challenging. Most time series analysis tools are limited
for the analysis of this kind of data. Recurrence analysis, a powerful concept from
nonlinear time series analysis, provides several opportunities to work with 
event data and even for the most challenging task of comparing event time series
with continuous time series. Here, the basic concept is introduced, the challenges
are discussed, and the future perspectives are summarised.

\tiny
 \keyFont{ \section{Keywords:} event time series, extreme events, recurrence analysis, edit distance, synchronisation} 
\end{abstract}

\section{Introduction}

The study of event time series is of general interest in data analysis and modelling,
because of their ubiquitous nature in almost all scientific fields, such as
investigating financial transactions, customer interactions, life-threatening cardiac events,
system failures, or natural phenomena. Event series can be single, discrete events, 
binary events or events with different amplitude, e.g., events extracted from data
with heavy tail distributions, short-term extreme events, or anomalies in time series.
In neuroscience, event series are also called ``spike trains'' \citep{harris2002}.
A time series is generally be denoted by a set of ordered pairs $\{(t_i, x_i)\}$ of time $t_i$ with $t_{i+1} > t_{i}$ 
and corresponding data value $x_i$; and with sampling
index $i$ (usually constant sampling time $t_{i+1} - t_{i} = \text{const.}$, 
i.e., equidistant time axis). 
An event series, instead, is considered as a series of event times,
defined by the associated specific time or timestamp of the single events, 
finally resulting in a set of event time points $\{t_i\}$.
As events could also have some amplitude, a definition as an event time series 
as a tuple of time and event strength $\{(t_i, x_i)\}$ is also possible.
Because the events usually do not occur at 
regular intervals, such event time series are usually irregularly sampled
$t_{i+1} - t_{i} \ne \text{const.}$
The alternative is using a regularly sampled, 
discretised time axis with binary (or amplitude) values at those points of time
where the event happens (this is similar to categorical data, another class
of discrete data, but not necessarily representing separated single 
events). 
However, this approach is usually limited and not appropriate
for many research questions, 
because the timing of events often does not fit the sampling points and, even more important, the time
series can be filled with many zeros. Standard time series analysis tools
have their limits when analysing such data. 

Examples of event data are time series of extreme events,
which are of specific interest because of their technical and medical importance, 
and their potential of serious societal impacts: Extreme rainfall (flush floods) 
and river floods are of 
strong concern because they are increasing
worldwide due to the global warming \citep{prein2017,kemter2020}. 
Extreme loading conditions are considered
and modelled in material sciences to monitor and predict serious failures, e.g., on bridges 
caused by extreme traffic or on airplane structures due to sudden stress or 
birdstrikes \citep{enright2013,iannucci2006}.
Ventricular tachyarrhythmias are life-threatening cardiac arrhythmias, usually analysed by
investigating the beat-to-beat intervals of the heart, expressed by a series of events 
(described as the R-waves in an electrocardiogram (ECG)), themselves \citep{marwan2002herz}.
Examples of natural, rare event time series are sequences of landslides. 
Such landslide events cause serious damages and can be triggered by specific weather phenomena, like
atmospheric rivers or El Ni\~no/ Southern Oscillation \citep{miller2019,sepulveda2015}. They are
increasingly affecting urban settlements, because of the spreading
of cities and climate change \citep{ozturk2022}.
Another example is brain activity which is controlled by the firing of neurons.
For example, the coherence of neuron firings can cause seizures \citep{steriade2000}.
The investigation of extreme events in dynamical systems is an important
subject in statistics and statistical physics. It covers many research questions, from
the emergence of extremes to predicting extreme events \citep{nagchowdhury2022}.

The research questions related to event series are often the same as for other kind of data,
e.g., comparing different time series, classifying the dynamics of the process behind, 
identifying regime changes, or use as the base for simulations and predictions.
The growing availability of data and computational facilities in almost all 
scientific disciplines has significantly advanced data science in general.
Several approaches have been introduced that allow to study
different research questions related of event data 
\citep{malik2012,ciba2020,voit2022}. Among them are probabilistic methods
based on large deviation and extreme value theory 
(parametric, semi-parametric approaches,
multivariate extensions), pattern-based prediction algorithms and
BDE modeling \citep{beirlant2006,ghil2011,lucarini2016}, 
as well as modern learning based approaches for predicting extreme events
\citep{qi2020,banerjee2022}.
Another class of methods are based on
the property of recurrences of states. In general, recurrence
based methods provide versatile approaches for classifying data, identification
of regime transitions, but also for unveiling interrelationships, synchronisation,
and causal links between different dynamical systems \citep{eckmann87,romano2004,hirata2010,marwan2023}. 
Due to its broad usability, simplicity, and growing number of software allowing recurrence analysis \citep{rpwebsite_software},
this method is attracting more and more interest and becoming increasingly popular \citep{marwan2008epjst,marwan2023}.
By modifying the definition of what a recurrence is, it is, in general, possible to adapt 
recurrence analysis to be usable for analysing discrete and event-like data \citep{faure2010,suzuki2010,banerjee2021}.
In the following, I describe briefly the basics of recurrence analysis,
its extension to work with event data, and the related challenges and future perspectives.

\section{Recurrence analysis of event time series}

A recurrence plot (RP) is the graphical representation of the recurrence matrix, which is 
simply representing all pair-wise time combinations $(i,j)$ of a data sequence which have similar values
or states $\vec{x}_i$:
\begin{equation}\label{eq_rp}
R_{i,j} = \Theta\left(\varepsilon - d(\vec{x_i}, \vec{x_j})\right),
\end{equation}
with a similarity measure $d(\cdot, \cdot)$ and the Heaviside function $\Theta(\cdot)$ 
which sets $R_{i,j} = 1$ if the similarity value $d(\cdot, \cdot)$ falls below the
threshold $\varepsilon$ \citep{marwan2007}. For dynamical systems with continuous change of the state variables, 
i.e., $\vec{x}_i \in \mathbb{R}^m$ (with $m$ the dimension of the system), the similarity
between states is often defined by the Euclidean norm 
$d(\vec{x_i}, \vec{x_j}) = \left\|\vec{x}_i - \vec{x}_j \right\|$ \citep{marwan2007}.
For discrete data of regular sampling (equidistant time instances), e.g., categorical data, the
recurrence matrix $\mathbf{R}$ can be simply defined by the Kronecker delta
function $R_{i,j} = \delta(x_i, x_j)$, which is one if both arguments are identical
\citep{groth2005,bandt2008,faure2010,leonardi2018}.
This approach works well for discrete data, such as categorical data or symbolic sequences,
with applications, e.g., in life science to detect atrial fibrillation or congestive heart failure \citep{caballeropintado2018,perezvalero2020b}, to measure synchronisation in an
epileptic brain \citep{groth2005}, or in engineering to optimise manufacturing networks \citep{donner2008}.
This concept is easily extendable for bivariate analysis. Cross-RPs, $CR_{i,j}^{x,y} = \Theta\left(\varepsilon - d(\vec{x_i}, \vec{y_j})\right)$, and joint-RPs, 
$JR_{i,j}^{x,y} = \Theta\left(\varepsilon - d(\vec{x_i}, \vec{x_j})\right) \circ \Theta\left(\varepsilon - d(\vec{y_i}, \vec{y_j})\right)$, are two 
basic concepts for measuring different aspects of synchronisation \citep{marwan2007}. 
To modify, the cross-RP for discrete data, we can simply use the Kronecker delta
$CR_{i,j} = \delta(x_i, y_j)$ \citep{lirapalma2018}. Joint-RP even allows us to measure the
synchronisation between different types of data, such as discrete and continuous
data \citep{kodama2021}, where 
\begin{equation}\label{eq_JRP}
JR_{i,j}^{x,y} = \delta(x_i, x_j)  \circ \Theta\left(\varepsilon - d(\vec{y_i}, \vec{y_j})\right)
\end{equation}
is the Hadamard product of the RP of the discrete system $x_i$ and the
RP of the continuous system $\vec{y}_i$.

\begin{figure}[htbp]
\begin{center}
   \includegraphics[width=\columnwidth]{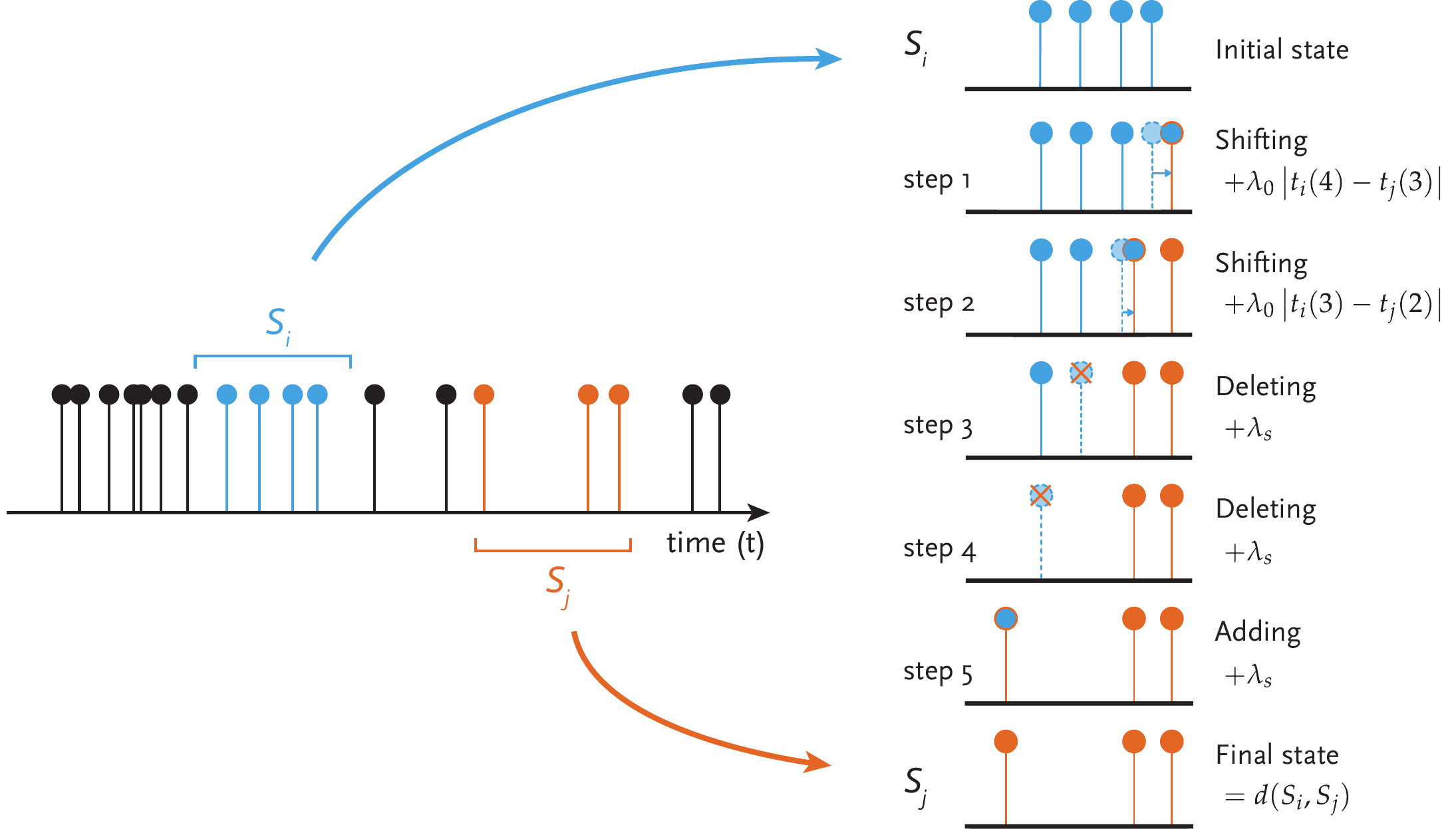}
   \caption{Edit distance as cost-based similarity between event sequences
   $S_i$ and $S_j$ from an event series (left).
   Events can be shifted, added or deleted, and their amplitude adjusted. All
   these operations have costs. The minimum cost is used as the
   distance (right).
   }\label{fig_editdistance}
\end{center}
\end{figure}

This concept reaches its limits when considering event time series which consist 
of rare events and many zeros between them, or, even more limiting, consist only of the 
events $\{t_i\}$ or have strong non-equidistant time instances ($t_{i+1} - t_{i} \ne \text{const.}$). For this kind of data,
the similarity measure $d(\cdot, \cdot)$ has to be replaced by a specific metric which
measures the coincidence of event sequences. Several measures (event metrics) are 
available, mainly developed in neuroscience \citep{ciba2020}. A widespread measure 
would be the event synchronisation, which allows varying delays between events
to be considered as coinciding \citep{quiroga2002}. This measure is successfully
applied for investigating, e.g., the spatio-temporal relationships between extreme 
rainfall events \citep{boers2016}. Another candidate is the edit distance,
an extension of the Levenshtein distance \citep{masek1980faster,victor1997}.
The distance is calculated by the minimum cost needed to modify one event sequence
into another with a limited set of operations (Fig.~\ref{fig_editdistance}).
Edit distance is a metric and has been successfully integrated with recurrence
analysis \citep{suzuki2010,banerjee2021}. 

The edit distance measure is the (minimum) sum of the costs of the transform
operations addition, deletion, and shifting applied to modify a sequence 
$S_i = \{t_1^{(i)}, t_2^{(i)}, \ldots, t^{(i)}_{N_i}\}$ of $N_i$ events
(with events at time points $t^{(i)}_a$) 
into sequence $S_j= \{t_1^{(j)}, t_2^{(j)}, \ldots, t^{(j)}_{N_j}\}$ (with events at time points $t_{b}^{(j)}$):
\begin{equation}\label{eq_ed}
d(S_i,S_j) = \min \Big\{ \lambda_s( N_i + N_j - 2N_{(i,j)} )+
\sum_{a, b \in \mathcal{C}}\lambda_0\| 
t^{(i)}_a-t^{(j)}_b\|\Big\},
\end{equation}
where $a$ and $b$ are indices of the events in segments $S_i$ and $S_j$; 
$N_i$ and $N_j$ the number of events in segments $S_i$ and $S_j$,
respectively; $N_{(i,j)}$ the number of events in $S_i$ and $S_j$ to be shifted,
which form the set $\mathcal{C}$; $\lambda_s$ is the cost of 
deletion/ insertion, and $\lambda_0$ the cost assigned for shifting events in time.
Thus, the first summand corresponds to deletion and insertion operations and the second summand to
the shifting of the events (Fig.~\ref{fig_editdistance}). Extensions of this cost function include considering
costs for amplitude changes or to modify the shifting term by a continuous cost function
allowing a more intuitive interpretation in terms of a delay \citep{suzuki2010,banerjee2021}.
To apply the edit distance for recurrence analysis, the event series has to be divided
into sequences $S_i$ defined by a time window of length $T_w$. The shifting of this
interval can be with smaller steps $s < T_w$ resulting in overlapping time intervals.
In order to get reliable costs $d(S_i,S_j)$, the resulting sequences $S_i$ and $S_j$ should
have at least one event (i.e., should not be empty).

\begin{figure}[htbp]
\begin{center}
   \includegraphics[width=\columnwidth]{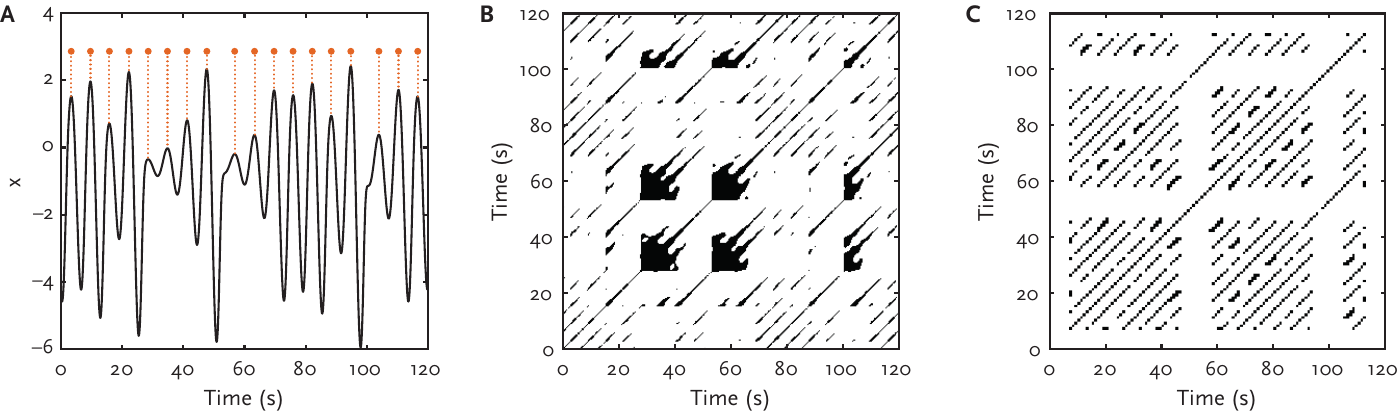}
   \caption{Example of a recurrence plot using edit distance. 
   (A) The maxima (red dots) of the $x$-variable of the R\"ossler system \citep{roessler1976}
   are used to mimic sparse (or extreme) events.
   (B) Recurrence plot calculated from the $(x, y, z)$-variables of the R\"ossler system
   (C) Recurrence plot derived from the events in (A) using the edit distance as defined
   by Eq.~(\ref{eq_ed}).
   Periodical occurrence of the events are clearly indicated by the period
   line structures in the edit distance recurrence plot. The empty bars around
   $t=$\SI{55}{\second} and $t=$\SI{100}{\second} indicate the parts in the dynamics with abrupt
   changes where no maximum values appear. Edit distance is calculated
   using overlapping windows length $T_w=$\SI{15}{\second} and moving step of $s=$\SI{1}{\second}. The recurrence
   threshold $\varepsilon$ is selected to ensure a recurrence rate (recurrence
   point density) of 15\%.}\label{fig_rp_roessler}
\end{center}
\end{figure}

This edit distance measure has been used as a simple synchronisation measure between event series
to study the stimulus responses in neuron spike trains \citep{victor1997},
as a similarity measure between extreme rainfall data to reconstruct climate networks \citep{agarwal2022},
and to create regularly sampled time series from non-regularly sampled time series (TACTS 
approach) \citep{ozken2015,eroglu2016}. It was also used as a distance measure
for computing RPs directly from event data (Fig.~\ref{fig_rp_roessler}), e.g., to study
stock exchange data \citep{suzuki2010}, flood events \citep{banerjee2021},
or to allow calculation of RPs directly from irregularly sampled palaeoclimate data \citep{ozken2018,ozdes2022}.
The integration of the edit distance metric into the RP definition, Eq.~(\ref{eq_rp}),
provides all the applications of recurrence based time series analysis for the 
specific data of event time series.

\section{Challenges}

Despite the recent advances in recurrence analysis of event time series, there are still several challenges.

Event time series can have missing data which are not easy to be detectable. 
For example, data on landslide events is mainly available at sites where they
affect infrastructure \citep{steger2021}, but their statistical analysis 
with respect to, e.g., climate change would require reliable
event series \citep{alvioli2018}.
Missing or sparse data can, therefore, bias the results of any analysis, and
is subject of research in time series analysis in general, 
including interpolation, modelling, or advanced data reconstruction 
methods \citep{alavi2006,facchini2011,sarafanov2022}, but mainly
not applicable for event data.

The process behind the analysed study object could be non-stationary
(e.g., life-threatening cardiac arrhythmias or seizures 
\citep{marwan2002herz,steriade2000}), meaning 
that the statistical properties of the event series may change over time (such as the 
distribution of events could change over time -- events may be sparse, 
meaning that there could be some periods of time without events), which can make it difficult to apply 
the event based recurrence analysis (e.g., using edit distance). 
In particular, if the time interval defined by 
length $T_w$ is too small, many sequences $S_i$ could be empty, resulting in non-defined costs 
$d(S_i,S_j)$. The selection of the time interval length $T_w$ is, thus,
crucial. For simple periodically recurring events, the choice might be easy,
but its selection if multiple time scales are present is not straightforward \citep{banerjee2021}.

The number of events in an interval can also change due to sampling issues, as it 
is a common problem in palaeoclimate research, where the deposition rate
in sediments is varying over time, thus, leading to palaeoclimate time series of
non-equidistant sampling in general \citep{rehfeld2011,breitenbach2012,braun2022}.
Event based metrics, such as event synchronisation, event coincidence analysis, or edit distance cost depend on the number of events in the interval 
and produce different types of biases which impact the results
of the quantitative analysis and
call for correction schemes \citep{wolf2020,braun2022}.

In general, the comparison of event time series with continuous time series is very
challenging. Such problems occur, e.g., in climate research when studying
recurring pattern of special weather phenomena (e.g., atmospheric rivers) or
extreme events (such as heavy precipitation or river floods)
with respect to large scale climate phenomena, such as El Ni\~no/ Southern Oscillation
or North Atlantic Oscillation \citep{miller2019,mundhenk2018}. 
The RP approach offers a promising way by modifying Eq.~(\ref{eq_JRP}) 
to
\begin{equation}\label{eq_JRP2}
JR_{i,j}^{x,y} = d(S_i^{x}, S_j^{x})  \circ \Theta\left(\varepsilon - d(\vec{y_i}, \vec{y_j})\right)
\end{equation}
 \citep{kodama2021}. However, event series
often consist of much less events than the number of sampling points of the continuous
time series, resulting in RPs of rather different length and making it impossible to
directly apply Eq.~(\ref{eq_JRP2}). An approach to finally match 
the event based RP with those of the continuous data would be required.

Finally, the uncertainty of the timing of events (timing jitter) is strongly affecting any measure
of coincidence. It is expected that timing jitter is a common problem
in measuring real-world event series. This challenge might be addressed 
by evaluating the sensitivity of the results on the jitter using specific
modells.

The extension of the edit distance can also take amplitude variations into account.
However, this mixes two different aspects of the data: the temporal pattern
of event sequences and amplitude differences. The optimal choice of the 
corresponding parameters might be less clear then, but have to be used to
balance between these aspects.

\section{Discussion}

The perspective future methodical developments will consider several important
challenges to study interesting research questions related to (discrete) event data.

For recurrence analysis of event data, so far only the edit distance metric has been applied.
It would be important to test and compare also other measures, such as Needleman-Wunsch distance,
event synchronisation, event coincidence analysis, or ARI-SPIKE-distance \citep{needleman1970,quiroga2002,wolf2020,ciba2020}. Specific discrete
data might call for distance metrics considering amplitude differences,
e.g., edit distance or longest common subsequence \citep{bergroth2000}.

Data with missing events is a general problem. Different strategies might be considered
to solve this challenge, including correction and gap filling schemes \citep{braun2022,facchini2011}. Correction schemes are also important for
data with non-stationarities (varying sparsity of events). Such correction
schemes needs further development to be applicable in a more general 
way (e.g., independent of the event distribution) and be more computationally
efficient.

Events can exhibit some kind of temporal dependencies, meaning that the likelihood of an event 
may depend on the occurrence of previous events. The RQA measures could be used to study 
temporal dependencies from event series \citep{banerjee2021}. 
In general, diagonal lines in a RP represent the tendency that current neighbours in phase space 
will remain to be neighbours in the near future, thus represent serial dependence. The RQA
measure {\it determinism} is quantifying the fraction of recurrences forming such diagonal
lines and can, thus, be used as an indicator of serial dependence.

The classification of dynamical processes by event series based on duration, 
frequency of events or their characteristics (e.g., shape), will be another interesting
application which will also involve machine learning approaches. The combination
of machine learning with recurrence analysis is currently a strongly developing
field with applications mainly in classification and prediction,
using RPs and RQA measures as inputs in machine learning workflows \citep{marwan2023}. 
A typical example is to convert time series into images by using the RP approach
which are finally fed into the machine learning workflow for classification \citep{estebsari2020}.
RPs of event series can be used in a similar manner for such kind of classification
tasks. Other characteristics of event series (like serial dependence) would be accessible
to machine learning approaches by the RQA measures \citep{mohebbi2011,malekzadeh2021,yang2018b} .

The detection of interdependencies or synchronisation of (sub-)systems represented by 
different kinds of data (e.g., event data with continuous time series)
is an important methodical challenge. New approaches based on RPs seem to be promising,
including the concept of joint-RPs \citep{kodama2021} and the comparison of the probability of recurrences \citep{nkomidio2022}. The advantage is the comparison by the recurrence structure, which would
allow comparing time series of different kinds (e.g., event series vs.~continuous data).
It includes further developments to finally match the size of
event based RPs with those of the continuous data, e.g., considering
coarse-graining, interpolation, or specific window selections schemes (for 
event-sequence based metrics like edit distance) \citep{banerjee2021}.

RP based analysis can be used to infer coupling directions or even causal links between
different systems \citep{ramos2017,peluso2020}. Thus, the next step would be to
test this approach for its potential on causality testing even for event data.

RPs also allow to identify patterns or regularities in challenging data, such as event series, including 
the estimation of the power spectral density of event series \citep{kraemer2022b}.
The most obvious way to derive a spectrum from a RP is to use the probability of recurrence 
after lag $\tau$, which is simply the density of recurrence points along the diagonals (with distance
$\tau$ from the main diagonal). This probability of recurrence is related to the auto-correlation \citep{marwan2002pla,zbilut2008a}. Using the edit distance measure, the temporal 
dependency structure within the event series can be visualised and quantified with this approach.
Finally, the power spectrum can then be estimated from this probability of recurrence, either by
applying the Fourier transform or any other advanced decomposition \citep{zbilut2008a,kraemer2022b}.

The uncertainty of the timing of events (timing jitter) needs to be considered in the analysis, leading
to new concepts such as Monte Carlo based ensemble approaches or Bayesian approaches.
A recently proposed concept combines a Bayesian approach
with RPs to derive a RP which explicitly represents the uncertainties of the 
timing of data points \citep{goswami2018}.
The resulting recurrence matrix contains the {probabilities of recurrences} 
instead of the binary information of recurrences. The recurrence quantification of such
matrix is still subject of future research.

Although various distance measures for event based RP computation are available,
the already applied one, edit distance, provides already a bunch of interesting
directions for future research. For example, the choice of an optimal
window length $T_w$ or the different cost parameters $\lambda$.
Including the cost for amplitude differences require an optimal choice of the 
corresponding parameters, which would need some systematic studies to provide
some guidance to balance between the differences in the temporal and spatial
domain.

The recurrence analysis as a concept is rather novel approach, with a lot of interesting and powerful
developments and extensions in the last two decades \citep{marwan2023}. It is
also a promising concept for studying different aspects related to (discrete) event time series,
where other methods have their limits.

\section*{Conflict of Interest Statement}

The author declares that the research was conducted in the absence of any commercial or financial relationships that could be construed as a potential conflict of interest.

\section*{Author Contributions}

The author has conceived the work and has written the manuscript.

\section*{Code availablity}

Julia code to reproduce Fig.~\ref{fig_rp_roessler} is available at Zenodo
\url{https:doi.org/10.5281/zenodo.7467886}.

\section*{Funding}
This work has been partly supported by the BMBF grant climXtreme (No.~01LP1902J) ``Spatial 
synchronization patterns of heavy precipitation events'' and by the DFG research training 
group GRK 2043/1 ``Natural risk in a changing world (NatRiskChange)''.

\section*{Acknowledgments}
Tobias Braun, K.~Hauke Kraemer, Abhirup Banerjee, Deniz Eroglu, {\c C}elik \"Ozdes, and J\"urgen Kurths
are acknowledged for fruitful discussions and ongoing collaborations on this subject.

\bibliographystyle{Frontiers-Vancouver} 
\bibliography{manuscript}


\end{document}